\documentclass[amsmath,amssymb,reprint,superscriptaddress,nopacs,aps, prapplied,footnotebib]{revtex4-1}
\usepackage{graphicx,xcolor,float,hhline,braket}
\definecolor{red}{rgb}{1,0.,0}
\newcommand{\fix}[1]{{#1}}
\usepackage[absolute]{textpos}

\begin{document}

%%% TITLE et al. %%%%
\title{Quantum Secret Sharing with Polarization-Entangled Photon Pairs}
\author{Brian P. Williams}
\email{williamsbp@ornl.gov}
\affiliation{Quantum Information Science Group, Computational Sciences and Engineering Division, Oak Ridge National Laboratory, Oak Ridge, Tennessee USA 37831}
\author{Joseph M. Lukens}
\affiliation{Quantum Information Science Group, Computational Sciences and Engineering Division, Oak Ridge National Laboratory, Oak Ridge, Tennessee USA 37831}
%\email{lukensjm@ornl.gov}
\author{Nicholas A. Peters}
\affiliation{Quantum Information Science Group, Computational Sciences and Engineering Division, Oak Ridge National Laboratory, Oak Ridge, Tennessee USA 37831}
\affiliation{Bredesen Center for Interdisciplinary Research and Graduate Education, The University of Tennessee, Knoxville, Tennessee 37996, USA}
\author{Bing Qi}
\affiliation{Quantum Information Science Group, Computational Sciences and Engineering Division, Oak Ridge National Laboratory, Oak Ridge, Tennessee USA 37831}
\affiliation{Department of Physics and Astronomy, The University of Tennessee, Knoxville, Tennessee 37996, USA}
%\email{petersna@ornl.gov}
\author{Warren P. Grice}
\altaffiliation[Current address: ]{Qubitekk, LLC, Vista, California 92081, USA}
\affiliation{Quantum Information Science Group, Computational Sciences and Engineering Division, Oak Ridge National Laboratory, Oak Ridge, Tennessee USA 37831}

\begin{abstract}
We describe and experimentally demonstrate a \fix{more practical} three-party quantum secret sharing \fix{(QSS)} protocol using polarization-entangled photon pairs. The source itself serves as an active participant and can switch between the required photon states by modulating the pump beam only, thereby making the protocol less susceptible to loss and amenable to fast switching. \fix{We derive a security proof based on quantum key distribution, demonstrating our QSS protocol to be secure against both eavesdropping and dishonest participants.} Compared to three-photon protocols, the practical efficiency is dramatically improved as there is no need to generate, transmit, or detect a third photon.   
\end{abstract}
%\pacs{03.67.-a, 03.67.Bg,03.67.Dd,03.67.Hk,42.50.Dv}
%03.67.-a Quantum information
%03.67.Bg Entanglement production and manipulation
%03.67.Dd Quantum cryptography and communication security
%03.67.Hk Quantum communication
%42.50.Dv Quantum state engineering and measurements

\maketitle

%\begin{textblock}{13.45}(1.35,14.9)
%\noindent \fontsize{7}{7}\selectfont This manuscript has been co-authored by UT-Battelle, LLC, under contract DE-AC05-00OR22725 with the US Department of Energy (DOE). The US government retains and the publisher, by accepting the article for publication, acknowledges that the US government retains a nonexclusive, paid-up, irrevocable, worldwide license to publish or reproduce the published form of this manuscript, or allow others to do so, for US government purposes. DOE will provide public access to these results of federally sponsored research in accordance with the DOE Public Access Plan (http://energy.gov/downloads/doe-public-access-plan).
%\end{textblock}

\section{Introduction}%In quantum key distribution (QKD), quantum correlations in states shared between two parties are used to establish a s
Most quantum security protocols, such as quantum key distribution (QKD)~\cite{Bennett1984, Ekert1991, Gisin2002, Lo2014}, are designed for two parties, %While appropriate for applications such as encrypted one-to-one communication,
yet many practical security situations involve multiple parties. An example is secret sharing, in which a secret distributed to members of a group can be reconstructed only when a sufficient number of the group members combine their respective portions~\cite{Blakley1979, Shamir1979}. In \emph{quantum} secret sharing (QSS), the partial secrets are distributed to $N$ participants via quantum states in such a way that any subset of them containing at least $k$ parties (with $k\leq N$) can combine their information to determine the information possessed by the dealer. As initially  envisioned~\cite{hillery1999quantum,Xiao2004, Chen2005}, QSS relies on distributing an $(N$$+$$1)$-partite Greenberger--Horne--Zeilinger (GHZ) state~\cite{Greenberger1989} to $N$$+$$1$ users who perform measurements; any one party can function as the dealer, or secret holder, and the remaining $N$ players can determine the dealer's result only if they collaborate, making this a QSS technique for disseminating shared \emph{classical} information. Other protocols have focused on QSS of \emph{quantum} information~\cite{Cleve1999, Gottesman2000}, and a variety of alternative multipartite entangled resources have been considered in both  discrete-~\cite{Deng2005, Gaertner2007, Markham2008, Keet2010, Bell2014, Lu2016} and continuous-variable~\cite{Tyc2002, Lance2003, Lance2004, Kogias17, Zhou2018} encodings.  Yet large quantum states are extremely challenging to realize in the laboratory, and more practical QSS versions have emerged, based on simpler quantum resources such as entangled photon pairs~\cite{Karlsson1999, Tittel2001, Chen2007, Grice2018} or single photons~\cite{schmid2005experimental, Bogdanski2008, han2008multiparty, Hai2013}.

In this Letter, we introduce a modified entangled-photon QSS protocol optimized for polarization qubits. In contrast to the original proposal for Bell pairs~\cite{Karlsson1999}, preparing input states in our case is possible by modulation of the pump beam only, so that fast and lossy \fix{phase} modulators can be incorporated \emph{before} quantum state generation, thereby preserving entanglement downstream. %The ``one-way'' nature of this three-party protocol is also immune to Trojan-horse attacks~\cite{gisin2006trojan}.
We experimentally observe the quantum correlations required for our protocol, obtaining, on average, three-party correlations at $(89.3\pm 0.5)\%$ with respect to their ideal values. Bolstered by a security proof built on QKD, our QSS protocol is practical and well-suited to current technology.%, and could find application in multiuser quantum communications networks.

\section{Background}
At a high level, QSS relies on correlations present in the preparation and measurement of some quantum system. For one-to-$N$-party QSS of classical information, each user (including the dealer) possesses two bits of data: one bit is \emph{public}, which is shared freely among all parties; the other is \emph{private}, revealed only to a subset of players working together to determine the private bit of the dealer (the secret). While both classical and quantum secret sharing can be considered for subsets of size $k$ ($k\leq N$) in so-called $(k,N)$ threshold schemes, here we focus on QSS for the particular case $k=N$; that is, only all $N$ players sharing their private bits can determine the private bit of the remaining $(N+1)^{\textrm{th}}$ party. %, whereas any combination of size $n<N$ has no knowledge of this bit, in the sense that each possibility is equally likely.
%For the original GHZ-based protocol~\cite{hillery1999quantum}, the quantum source is an $(N$$+$$1)$-qubit state of the form $\ket{\Psi} \propto \ket{00\cdots 0} + \ket{11\cdots 1}$; each party receives exactly one qubit from this state and performs a measurement in either the Pauli $X$ or $Y$ basis (the public bit), with the specific result (eigenvalue $+1$ or $-1$) serving as the private bit. For half of all public bit combinations, a secret is generated, where a size-$N$ collection of secret bits is necessary and sufficient to decode that of the dealer.

%The GHZ version

The original GHZ-based protocol~\cite{hillery1999quantum} enjoys a satisfying symmetry for all parties, with public bits being measurement \emph{bases} and private bits measurement \emph{results}. Yet GHZ states are notoriously difficult to generate, so that experimental QSS demonstrations rely on intrinsically rare events, such as two-pair creation in spontaneous parametric downconversion (SPDC)~\cite{Chen2005}. Fortunately, one-to-$N$ QSS does not actually require $(N+1)$-partite entanglement, since each party need not measure a quantum system---only modify it. For example, a single-photon protocol in which intermediate parties rotate the state via one of four unitaries produces the same correlations as measurements of a GHZ state~\cite{schmid2005experimental}. Likewise, a source of entangled photon pairs switching between four states permits QSS in which the source is an active participant~\cite{Karlsson1999}. While demonstrated experimentally for time-bin entangled photons~\cite{Tittel2001}, to our knowledge no implementation of two-photon QSS with polarization qubits has been realized.

\begin{figure}[t]
\includegraphics[width=\linewidth]{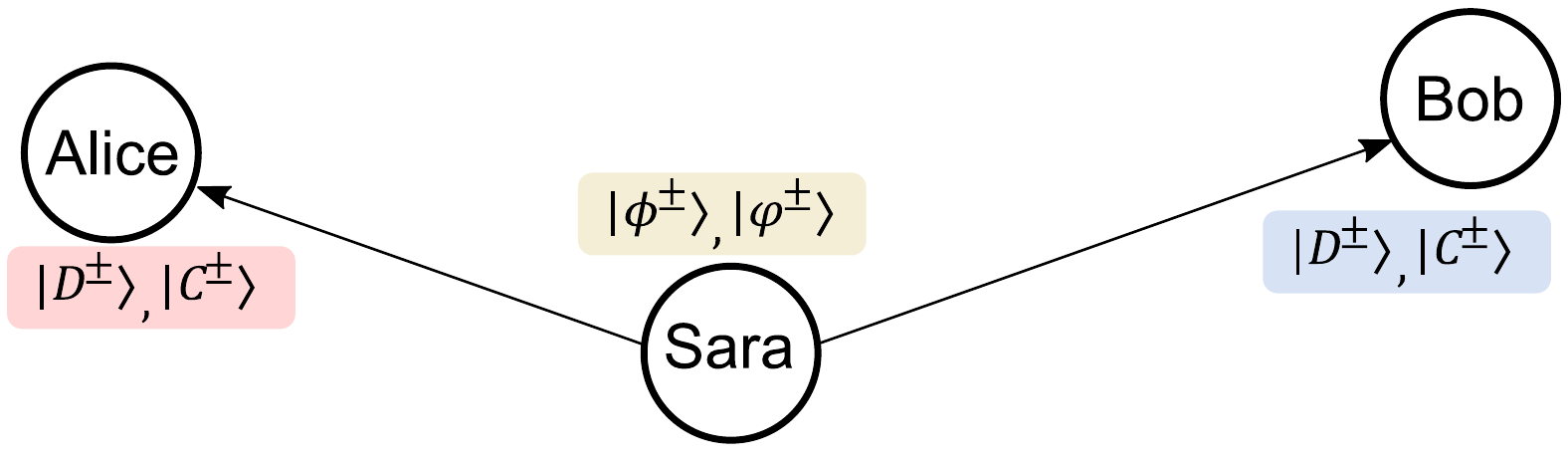}
\caption{\fix{Three-party quantum secret sharing protocol. Sara chooses one of four entangled states; Alice and Bob randomly select to measure in either the diagonal or circular bases, recording their measurement results.}}
\label{fig1}
\end{figure}

\section{Protocol}
Figure~\ref{fig1} outlines our proposed protocol. We consider the three-party correlations between source Sara and receivers Alice and Bob. Central to enabling a practical polarization-entangled version of QSS is Sara's choice of states. %For a total of two bits, she must select from at least four states, and an adversary must not be able to determine her private bit. Incidentally, this eliminates the four Bell states from consideration, for while satisfying the two-bit requirement, these fail to provide security: it would be possible, in principle, for an adversary to perform a Bell-state measurement and recover all of Sara's information deterministically~\cite{Karlsson1999}
We select the two $\phi$ Bell states as one basis, and ``plus/minus $i$'' states as the other:
\begin{equation}
\begin{aligned}
\ket{\phi^\pm}&=\tfrac{1}{\sqrt{2}}\left(\ket{HH}\pm\ket{VV}\right)\\
\ket{\varphi^{\pm}}&=\tfrac{1}{\sqrt{2}}\left(\ket{HH}\pm i \ket{VV}\right),
\label{saraStates}
\end{aligned}
\end{equation}
where $\ket{mn}$$\equiv$$\ket{m}_\mathrm{Alice}\otimes \ket{n}_\mathrm{Bob}$ and $\ket{H}$ ($\ket{V}$) denotes the horizontal (vertical) polarization eigenstate. Any measurement that would unambiguously identify the first two states would not be able to discern the second pair, and vice versa, due to the fact that $|\braket{\phi^\pm|\varphi^\pm}|^2 = |\braket{\phi^\pm|\varphi^\mp}|^2 = \tfrac{1}{2}$. In this way, we can define Sara's public bit $S$ as the basis choice $S\in\{\phi,\varphi\}$ and her private bit $s$ as the state within that basis, $s\in\{+,-\}$.%, for the top/bottom of $\pm$.

Importantly, these states are more convenient for experimental implementation than those considered in previous proposals for two-photon QSS, where linear combinations of Bell states form the second basis~\cite{Karlsson1999, Grice2018}. Use of a single source in that case for preparation of all four states requires polarization rotation of one of the two photons \emph{after} generation. On the other hand, all the states in Eq.~(\ref{saraStates}) share the same correlations in $H$ and $V$, differing only in relative phase between the $\ket{HH}$ and $\ket{VV}$ contributions. Switching between all states can be effected by modifying the phase between the $H$ and $V$ components of the pump beam before generation. 
Since loss experienced by the pump has no impact on state generation (aside from needing more pump power), one can employ high-speed and potentially lossy phase modulators without degrading the performance of QSS. 
%\fix{Moreover, Alice and Bob are free to use passive splitting to select their measurement bases, where they are informed of basis choice by photon detection, thereby permitting low-loss measurement as well--[\textit{Is this somehow immune to detector saturating attack? Maybe leave this part out?}]}.

Alice and Bob measure in either the diagonal ($D$) or circular ($C$) bases, with eigenstates
%\begin{equation}
%\begin{aligned}
%\ket{D^\pm}&=\tfrac{1}{\sqrt{2}}\left(\ket{H}\pm\ket{V}\right) \\
%\ket{C^\pm}&=\tfrac{1}{\sqrt{2}}\left(\ket{H}\pm i\ket{V}\right).
%\label{eigenstate}
%\end{aligned}
%\end{equation}
\begin{equation}
\ket{D^\pm}\!=\!\tfrac{1}{\sqrt{2}}\left(\ket{H}\pm\ket{V}\right);\quad
\ket{C^\pm}\!=\!\tfrac{1}{\sqrt{2}}\left(\ket{H}\pm i\ket{V}\right).
\label{eigenstate}\end{equation}
We can define Alice's public bit $A\in\{D,C\}$ (basis choice) and private bit $a\in\{+,-\}$ (result), and similarly for Bob: public bit $B\in\{D,C\}$ and private bit $b\in\{+,-\}$. There are four combinations of public bits $(S,A,B)$ that produce a secret, identified by expressing Eq.~(\ref{saraStates}) in terms of  $\ket{D^\pm}$ and $\ket{C^\pm}$:
\begin{equation}
\begin{aligned}
\ket{\phi^\pm}&=\tfrac{1}{\sqrt{2}}\left(\ket{D^+ D^\pm}+\ket{D^- D^\mp}\right)\\
&=\tfrac{1}{\sqrt{2}}\left(\ket{C^+ C^\mp}+\ket{C^- C^\pm}\right)\\
\ket{\varphi^\pm}&=\tfrac{1}{\sqrt{2}}\left(\ket{D^+ C^\pm}+\ket{D^- C^\mp }\right)\\
&=\tfrac{1}{\sqrt{2}}\left(\ket{C^+ D^\pm}+\ket{C^-D^\mp}\right). \label{corr}
\end{aligned}
\end{equation}
In other words, the correlations required for QSS appear when Alice and Bob choose the same basis for the $\phi$ states, and different bases for the $\varphi$ states. If Sara selects from $s\in\{+,-\}$ with equal probability, the private bits $(s,a,b)$ in each of the above lines assume all possibilities with equal probability individually and pairwise, yet collectively they are perfectly correlated. We can express the quantum correlation via the total parity (product of the signs), $\varepsilon = sab$, which adopts the value $+1$ $(-1)$ for an even (odd) number of negative private bits, and has a definite value in each of the rows in Eq.~(\ref{corr}).

Table~\ref{tab1} provides a summary of the public bit combinations leading to quantum correlations; for the other four public bit selections, the outcomes $\varepsilon =\pm 1$ are equally probable, and QSS is not possible. (As in other QSS protocols, this 50\% failure rate can be circumvented by selecting basis combinations asymmetrically~\cite{Xiao2004}.) %With this table, we briefly outline how the secret sharing protocol proceeds (similar in form to ~\cite{Karlsson1999}):

We now describe how our QSS protocol proceeds and outline its general security against both external eavesdroppers and dishonest participants.

\vspace{0.05in}
\textit{Quantum stage}
\vspace{-0.1in}
\begin{enumerate}
\setlength\itemsep{0pt}
\item Sara randomly selects public bits $S\in\lbrace \phi,\varphi \rbrace $ and private bits $s\in \lbrace +,-\rbrace $, prepares the state $\vert S^s\rangle$, and sends the photons to Alice and Bob.
\item Alice and Bob randomly and independently choose to measure their photons in either the $D$ or $C$ basis.
\item When both Alice and Bob detect photons, the three participants keep their data as raw key.
\item They repeat (1)-(3) to generate more raw key.
\end{enumerate}

After the quantum stage, one party is chosen as the dealer \fix{(either Sara, Alice, or Bob)} while the other two serve as players.

\vspace{0.05in}
\textit{Classical post-processing stage}
\vspace{-0.1in}
\begin{enumerate}
\setlength\itemsep{0pt}
\item The dealer assumes player 1 is dishonest and player 2 is honest (there is no point to QSS if \emph{both} players are dishonest).
\item The dealer randomly selects a subset of raw key and requests that player 1 announce both the public and private bits.
\item The dealer and player 2 estimate a lower bound secure key rate $R_1$ for two-party QKD, under the assumption that player 1 collaborates with eavesdroppers (see details below). 
\item They repeat steps (1)-(3) reversing the roles of the two players, giving the lower bound secure key rate $R_2$.
\item The dealer determines the secure key rate $R$ of the QSS protocol as the minimum of $R_1$ and $R_2$.
\item The dealer requests the players to announce their public bits for the remaining data~\footnote{We remark that the order in which the two players announce their public bits in step 6 does not matter. This can be seen from Table 1: whether an event will be kept or not is determined by three public bits. As long as the dealer is the last one to reveal his/her public bit, the dishonest player cannot change the probability of a particular event being selected or not. }. Using Table 1 and all the public bits from the three parties, the dealer announces the transmissions leading to correlated data. All parties keep the corresponding private bits as the sifted key.
\item The dealer generates the final QSS key from the sifted key using one-way classical post-processing as developed in QKD. Collaboratively, the two players can recover the QSS key using their private bits and error correction information from the dealer. Alone, each of them gains only an exponentially small amount of information about the QSS key.
\end{enumerate}

\section{Security}
The security analysis of QSS is typically more involved than that of QKD. A security proof of QSS against eavesdroppers in the channels and dishonest players has only appeared recently~\cite{Kogias17}, which introduced the key idea to treat measurements announced by the players as input or output from an uncharacterized device, while the dealer is assumed trusted. This allows them to apply the tools developed in one-sided device-independent QKD~\cite{Tomamichel11} in the security analysis of QSS. Here, we extend the above idea by applying the security proof of standard QKD with trusted devices, which can yield a better key rate in practice. This is based on the observation that at least one of the players in QSS is honest (although the dealer does not know who). By evaluating the potential secure key rate of QKD with each individual player (assuming all the other players are dishonest) and using the smallest one as the QSS key rate, security against collaborating attacks between the eavesdropper and any $N$$-$$1$ players can be guaranteed. %(More details can be found in a forthcoming work \cite{CVQSS}.)
   
Here we briefly outline how to evaluate the secure key rate given that at most one of the two players is dishonest. There are two different cases:

\textit{Case 1: Dishonest player controls two-photon source.} Sara is the dishonest player; let us assume the dealer is Alice. After the quantum stage, Alice randomly selects a subset of raw data and requests that Sara announce which entanglement states she prepared in those events. Given the information announced by Sara, the QKD process between Alice and Bob becomes conventional entanglement-based QKD with an untrusted entanglement source between two honest users. Its unconditional security has been well established~\cite{Koashi03, Ma07}.    

\textit{Case 2: Dishonest player controls one set of detectors.}
Let us deem Alice the dishonest player. To prove security, we introduce a \emph{virtual} three-photon GHZ-state-based QKD protocol equivalent to the actual two-photon protocol. Specifically, we assume Sara prepares the state $\ket{\Psi}=\frac{1}{\sqrt{2}}\left( \ket{HHH} + \ket{VVV}\right) $ so that, by Eqs.~(\ref{saraStates})-(\ref{corr}),
\begin{equation}
\begin{aligned}
\ket{\Psi}&=\tfrac{1}{\sqrt{2}}\left(\ket{ D^+}\ket{\phi^+}+\ket{D^-}\ket{\phi^-}\right)\\
&=\tfrac{1}{\sqrt{2}}\left(\ket{C^-}\ket{\varphi^+} + \ket{C^+}\ket{\varphi^-} \right)
\label{eq9}.
\end{aligned}
\end{equation}

If Sara measures one photon in the $D$ ($C$) basis, depending on her measurement result, the other two photons will be projected onto $\ket{\phi^+}$ or $\ket{\phi^-}$ ($\ket{\varphi^+}$ or $\ket{ \varphi^-}$), which are the states she prepares in the actual protocol. This implies the equivalence between this virtual protocol using GHZ states and the actual protocol. Since the measurements of different participants commute with each other, we can switch the order without changing the statistics of the measurement results. We can imagine that both Sara and Bob will keep their photons until Alice announces her measurement basis and results. In this picture, if Alice follows the protocol, she will project the other two photons onto one of the four entangled states in Eq.~(\ref{saraStates}). With the information announced by Alice, Sara and Bob will know which entanglement state has been prepared. The QKD process between Sara and Bob again becomes conventional entanglement-based QKD.

\begin{table}
\begin{ruledtabular}
\caption{Combinations of public bits leading to correlated events, along with ideal values for the private bit correlation (second column), those measured experimentally (third column), and QBER for pairwise combinations (fourth column).\label{tab1}}
\begin{tabular}{cccc}
Public Bits     & $\braket{\varepsilon}$   & $\braket{\varepsilon}$ & Pairwise \\ $(S,A,B)$  & (Ideal) &  (Experiment) & QBER [\%] \\ \hline\hline
$(\phi,D,D)$    & $+1$ & $+0.89\pm0.01$ & $5.6\pm0.7$ \\
$(\phi,C,C)$    & $-1$ & $-0.88\pm0.01$ & $6.1\pm0.7$ \\
$(\varphi,D,C)$ & $+1$ & $+0.901\pm0.009$ & $4.9\pm0.6$ \\
$(\varphi,C,D)$ & $+1$ & $+0.90\pm0.01$ & $5.1\pm0.7$ 
\end{tabular}
\end{ruledtabular}
\end{table}

\begin{figure}[b]
\includegraphics[width=\linewidth]{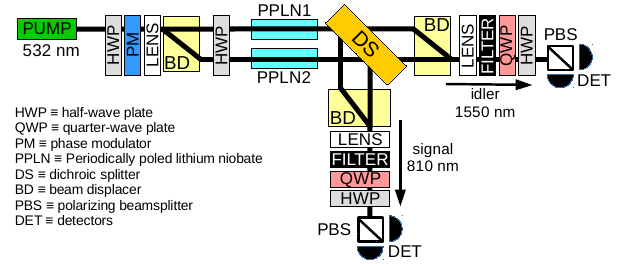}
\caption{Experimental setup. See text for details.
%Experimentally, using parallel 532 nm beams we pump a pair of nonlinear crystals, each of which is equally likely to output a signal-idler pair. The output paths are overlapped using beam displacers resulting in a polarization-entangled two-photon state.
}
\label{expResults}
\end{figure}

\begin{figure*}[!t]
\includegraphics[width=0.8\textwidth]{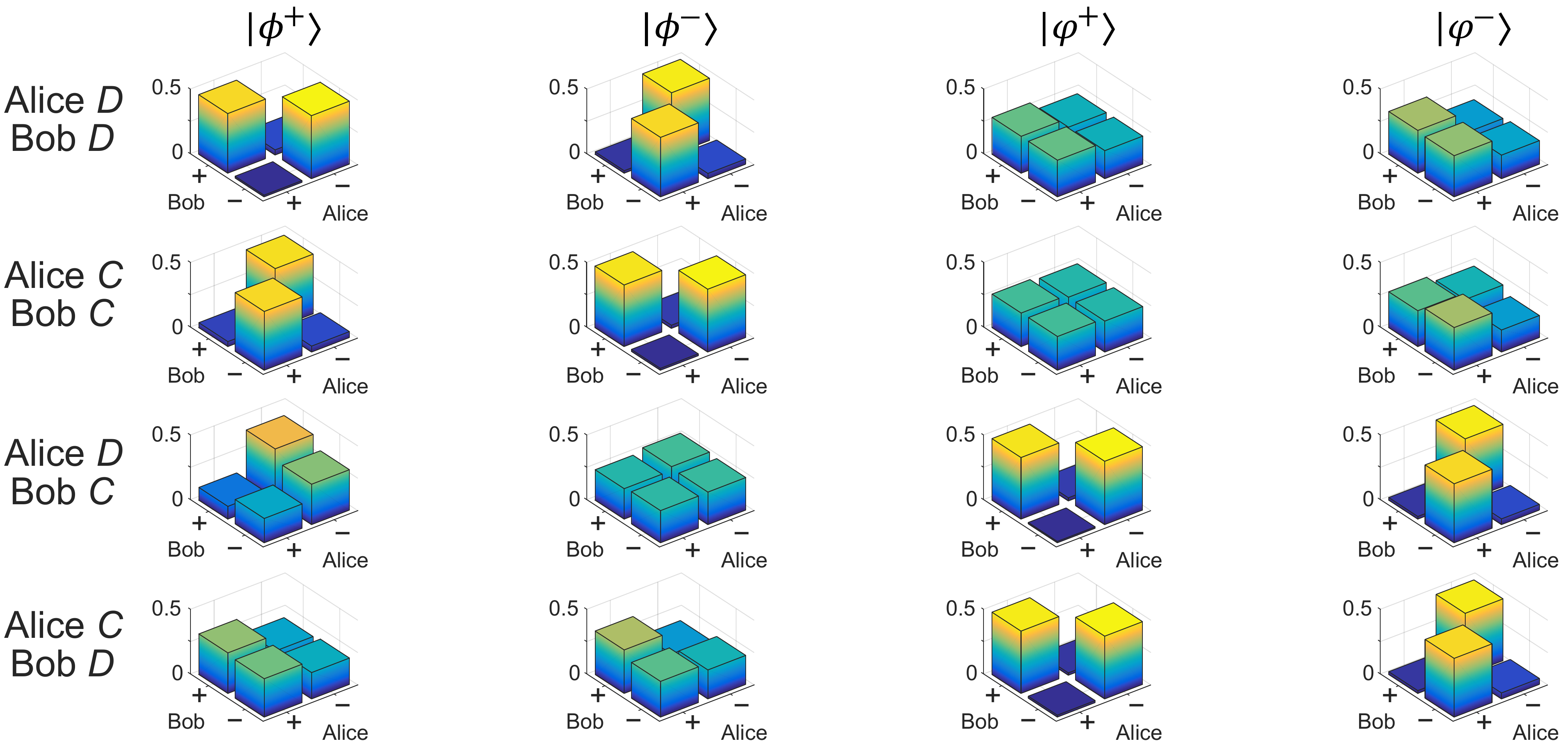}
\caption{Experimentally measured outcome probabilities for Alice and Bob, conditioned on input state and measurement bases.}
\label{QSSplots}
\end{figure*}

\section{Experiment}
In order to explore these correlations experimentally, we utilize the polarization-entangled photon source in Fig.~\ref{expResults}. Based on a similar design for 1550~nm entangled photons~\cite{evans2010bright}, this source relies on type-0 SPDC in two parallel, orthogonally rotated, periodically poled lithium niobate (PPLN) crystals. With a combination of beam displacers preceding and following the two crystals, we are able to generate a superposition of $HH$ and $VV$ photon pair amplitudes in a single spatial mode. This design is stable and has been shown to offer bright, high-fidelity entanglement~\cite{evans2010bright}, with a modified version applied in one of the seminal loophole-free Bell inequality tests~\cite{Shalm2015}. The main difference here is the creation of non-degenerete photons; we pump with a frequency-doubled Nd:YAG laser at 532~nm and produce photons at 810~nm (Alice) and 1550~nm (Bob), using a dichroic splitter to separate them. By modulating the phase between the two pump polarizations prior to the first beam displacer, we ideally produce the state $\ket{\Psi}\propto \ket{HH}+e^{i\alpha}\ket{VV}$, with the phase $\alpha$ tunable from 0 to $2\pi$, producing Sara's QSS states [Eq.~(\ref{saraStates})]. %$\ket{\phi^+} (\alpha=0)$, $\ket{\phi^-} (\alpha=\pi)$, $\ket{\varphi^+} (\alpha=\pi/2)$, and $\ket{\varphi^-} (\alpha=3\pi/2)$.

To characterize each of these states, we utilize polarization analyzers in both of the photon arms. To reduce noise on the InGaAs avalanche photodiodes (APDs) used for the 1550~nm photon, we trigger them with detections on the free-running Si APDs for the 810~nm photon using an FPGA. %, so that a click on one of the InGaAs APDs signifies a coincidence.
%Detection time-tagging and gate signal manipulation are enabled using an FPGA.
As an example, the fidelity for preparation of the state $\ket{\phi^+}$ we measure at $\braket{\phi^+ | \rho |\phi^+} = 0.949 \pm 0.001$ (without subtraction of accidentals), where we use Bayesian tomography to reconstruct the full two-photon density matrix $\rho$~\cite{Blume2010, Williams2017}.

To test all QSS correlations, we prepare Sara's four states and measure the coincidences in every $D/C$ basis combination for Alice and Bob. After backing out loss and detector efficiency~\cite{Williams2017}, we obtain the sixteen normalized probability distributions in Fig.~\ref{QSSplots}. Each column denotes a particular state, while each row shows a specific Alice/Bob basis combination. As expected, strong correlations result for the cases in Table~\ref{tab1}, while near-uniform probabilities are obtained for the remaining public bit combinations. We can quantify the correlation in each case using an Alice/Bob parity measure, $\varepsilon_s = ab$, subscript $s$ signifying that this is conditioned on a specific state from Sara. Its expectation value is given by
$\braket{\varepsilon_s} = P_{++} + P_{--} - P_{+-} - P_{-+}$,
with the probabilities taken from the relevant plot in Fig.~\ref{QSSplots}.

\begin{figure}[!ht]
\includegraphics[width=0.9\linewidth]{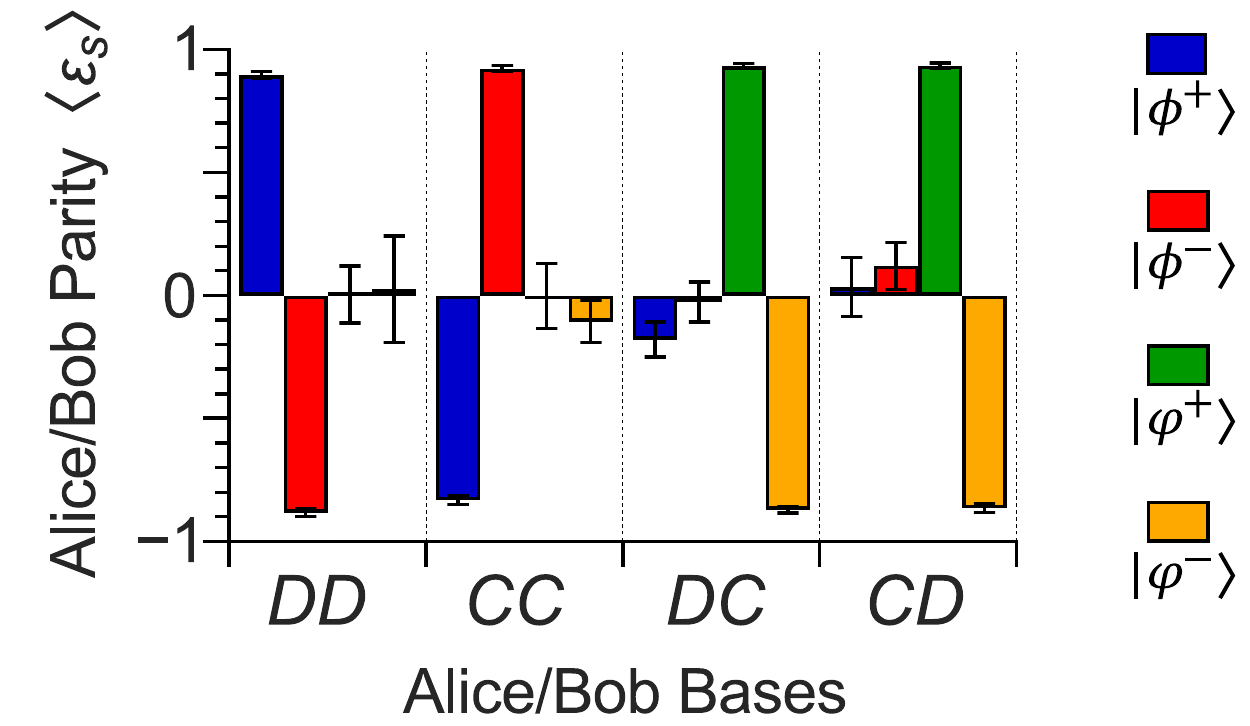}
\caption{Experimentally measured correlations, as expressed by expectation of the parity of Alice's and Bob's results.}
\label{QSScorr}
\end{figure}

We calculate all sixteen of these correlation measures and display them in Fig.~\ref{QSScorr}. (Error bars are $\pm$ one standard deviation, computed from the Bayesian posterior distribution underlying the probabilities in Fig.~\ref{QSSplots}.) For the basis combinations $DD$ and $CC$, we find strong correlations with the $\ket{\phi^\pm}$ states, while for $DC$ and $CD$ the correlations are strong for $\ket{\varphi^\pm}$. We can combine pairs of these results to compute the expectation of the three-party correlation parameter $\braket{\varepsilon} = \frac{1}{2} \left( \braket{\varepsilon_{s=+}} - \braket{\varepsilon_{s=-}} \right)$, where we have assumed that each of Sara's options $s\in\{+,-\}$ are chosen with equal probability. %, and the minus sign before the second term reflects the flip in parity of the quantity $\varepsilon=sab$ when $s$ is negative.
This yields the numbers in the third column of Table~\ref{tab1}, whose absolute values average to $0.893 \pm0.005$ compared to the ideal of unity. The corresponding pairwise QBER values (as needed for security analysis) follow in the fourth column. In our example, with completely honest players, the QBER is identical for each pair of users and equal to $\tfrac{1}{2}(1-|\braket{\varepsilon}|)$. All values fall well below the 11\% threshold of entanglement-based QKD using one-way communication~\cite{Koashi03, Ma07}, indicating the capacity for a positive secure QSS rate in our setup. 

\section{Discussion}
\fix{While we have focused on pure QSS here, our physical setup is also compatible with QKD between any two of the three parties. With no modifications to the quantum stage of the protocol (state preparation and measurement), QKD between two participants can be realized by requiring the remaining party to reveal both public and private bits. This amounts to  implementing steps (1)-(3) of the classical post-processing stage, but now for all raw key values, not just a random subset. Such an approach to QKD is especially flexible in that the communicating parties can be decided after all raw data has been collected. Additionally, by adding intermediate participants between the source and receivers (as in single-qubit QSS~\cite{schmid2005experimental}), the protocol here can be extended to more than three users as well. Already considered as an extension of single-photon QSS~\cite{grice2015two}, the use of entangled photons should benefit from the same advantages of entangled-photon QKD in the traditional point-to-point setting---namely, no need for decoy states and increased tolerance to loss~\cite{Lo2014}. However, adding users between the source and detectors will require Trojan-horse countermeasures. Whereas the ``one-way'' configuration of our basic three-user QSS protocol is naturally immune to Trojan-horse attacks~\cite{gisin2006trojan}, the many-user case is susceptible to an eavesdropper sending in and extracting probe light to read the phase shifts applied by intermediate parties. This is a known QSS vulnerability~\cite{grice2015two} that applies in extending our approach. Thus security measures beyond the proof introduced here would be essential for additional parties.}

\fix{Finally, one of the primary practical advantages of our polarization-entangled QSS protocol is the reliance on pump polarization modulation to select Sara's quantum state, rather than modulation \emph{after} photon-pair generation. When placed before the downconversion crystal as done here, any introduced modulator loss can be compensated simply by turning up the pump power after the modulator but before the downconversion crystal, preserving the ideal pair emission probability. On the other hand, were a fast polarization modulator placed in either the signal or idler paths after generation to set the state, any loss would reduce the rate of transmitted entangled pairs, and hence the QSS rate as well. In this case, turning up the pump power to compensate would introduce additional noise from increased multipair emission, so that the secure key rate would not be maintained. As an example, typical commercial fiber-pigtailed modulators impart $\sim$3~dB loss, corresponding to either a two-fold reduction in coincidences, or equivalently a four-fold increase in multipair probability for the same coincidence rate, when comparing post-generation modulation against the pump modulation of our scheme.}

In conclusion, we have introduced a pragmatic approach for QSS with polarization-entangled photons, with security based on that of QKD. Relying on pump phase modulation to prepare the necessary entangled states, our method tolerates high-speed and lossy phase modulators without degrading the performance of QSS. Our experimental implementation reveals the correlations necessary for three-party secret sharing, and extending to additional parties should be possible with minor modifications. Our approach could extend two-photon entanglement-based QKD to multiple users by promoting the entanglement source to a user and adopting a QSS protocol. This practical use of quantum resources should benefit commercial cryptography development and may inspire further improvements to QKD and QSS implementations.

\begin{acknowledgments}
This work was performed at Oak Ridge National Laboratory, operated by UT-Battelle for the U.S. Department of Energy under Contract No. DEAC05-00OR22725. Funding was provided by the  U.S.  Department of Energy under the Cybersecurity for Energy Delivery Systems (CEDS) program.
\end{acknowledgments}

\bibliographystyle{apsrev4-1}
\bibliography{QSS.bib}

\end{document}